\documentclass[a4,aps,showpacs,amsmath,floatfix,twocolumn]{revtex4}
\usepackage{graphicx}
\usepackage{dcolumn}
\usepackage{color}
\usepackage{amsmath}
\usepackage{here}
\usepackage[utf8x]{inputenc}
\usepackage{bm}
\usepackage{amssymb}
\usepackage{amsmath}

\usepackage{hyperref}
\hypersetup{breaklinks=true,
            colorlinks=false,
            pdfborder={0 0 0}}
\usepackage{bm}
\usepackage{braket}
\usepackage{MnSymbol}
\usepackage{graphicx}
\graphicspath{{figures/}}
\usepackage{dsfont}

\begin{document}
\title{Nonlinear Hall Acceleration and the Quantum
Rectification Sum Rule}

\author{O. Matsyshyn$^{1}$, I. Sodemann$^{1}$}

\affiliation{$^{1}$Max Planck Institute for the Physics of Complex Systems, Dresden 01187 , Germany
}

\begin{abstract}
Electrons moving in a Bloch band are known to acquire an anomalous Hall velocity proportional to the Berry curvature of the band which is responsible for the intrinsic linear Hall effect in materials with broken time-reversal symmetry. Here, we demonstrate that there is also an anomalous correction to the electron acceleration which is proportional to the Berry curvature dipole and is responsible for the Nonlinear Hall effect recently discovered in materials with broken inversion symmetry. This allows us to uncover a deeper meaning of the Berry curvature dipole as a nonlinear version of the Drude weight that serves as a measurable order parameter for broken inversion symmetry in metals. We also derive a quantum rectification sum rule in time reversal invariant materials by showing that the integral over frequency of the rectification conductivity depends solely on the Berry connection and not on the band energies. The intraband spectral weight of this sum rule is exhausted by the Berry curvature dipole Drude-like peak, and the interband weight is also entirely controlled by the Berry connection. This sum rule opens a door to search for alternative photovoltaic technologies based on the Berry geometry of bands. We also describe the rectification properties of Weyl semimetals which are a promising platform to investigate these effects.
\end{abstract}

\pacs{72.15.-v,72.20.My,73.43.-f,03.65.Vf}

\maketitle

\textit{\color{blue} Introduction}. The Berry phase has become a protagonist on our modern understanding of the motion of electrons~\cite{RevModPhys.82.1959} and in the classification of their phases~\cite{RevModPhys.82.3045,RevModPhys.83.1057}. The Berry curvature determines an anomalous electron velocity which gives rise to the intrinsic Hall effect in materials without time reversal symmetry~\cite{RevModPhys.82.1539}.
It has been recently predicted ~\cite{2009arXiv0904.1917D,PhysRevLett.105.026805,PhysRevLett.115.216806} and experimentally observed~\cite{2018arXiv180908744K,Ma2019,2019arXiv190202699S} that materials with time reversal symmetry can display a $\textit{nonlinear Hall effect}$. The nonlinear Hall conductivity characterizing this effect is a product of the Berry curvature dipole (BCD), an intrinsic quantum geometric property of the material ~\cite{PhysRevLett.115.216806}, and the scattering time, which is subject to sample quality variations. This poses the question of whether the BCD could be probed or defined in a more fundamental manner that is less intertwined with disorder effects~\cite{2018arXiv181206989K,2018arXiv181208162I,2018arXiv181208377D,2019arXiv190104467N,2019arXiv190700577X}. 

In this Letter, we propose that the BCD is indeed a fundamental property of metals that plays the role of a nonlinear Drude weight, namely, it measures an anomalous non-Newtonian Hall acceleration that scales with the square of the electric field and that is allowed in materials without inversion symmetry. We will call this correction to the electron acceleration the nonlinear Hall acceleration. The first section of our work will be devoted to demonstrating this result in the low energy single band limit. Subsequently, by developing unifying theory of the second order optical and transport phenomena of metals and insulators, we will demonstrate a quantum rectification sum rule (QRSR), according to which the rectification conductivity of time-reversal-invariant materials integrates to a quantity that is entirely controlled by the Berry connection, and the BCD exhausts completely its intraband weight, as depicted in Fig.1, and in analogy to how the conventional Drude weight is related to the Drude peak in the linear conductivity of metals. This suggest that measurements of the rectified current over a broad frequency range offers a systematic way to estimate the BCD that bypasses detailed knowledge of the scattering rate, which could help disentangle the more subtle disorder mediated corrections~\cite{2018arXiv181206989K,2018arXiv181208162I,2018arXiv181208377D,2019arXiv190104467N,2019arXiv190700577X}  present in current experiments~\cite{2018arXiv180908744K,Ma2019,2019arXiv190202699S}. The BCD therefore offers a solution to the long-standing problem of defining a measurable order parameter for broken inversion symmetry in metals, since the electric polarization, which is a natural order parameter for broken inversion symmetry in insulators, is generally illdefined in systems where charge can freely flow. 

\begin{figure}[t]
    \centering
    \includegraphics[width=0.45\textwidth]{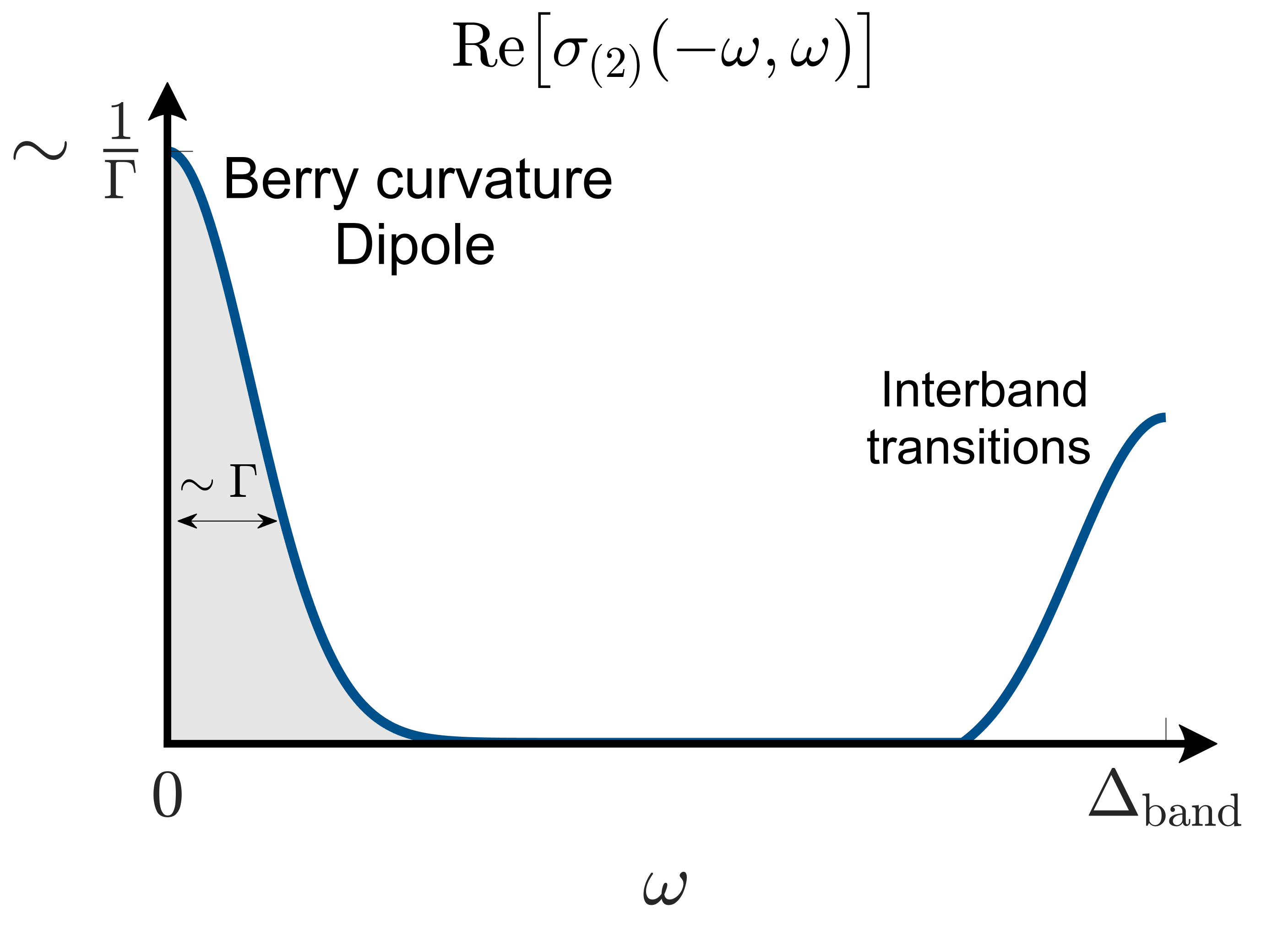}
    \caption{Illustration of the real part of the rectification conductivity which is the spectral weight on the QRSR in Eq.\eqref{sumR}. The intraband weight is exhausted by a Drude-like peak whose area equals the BCD up to universal constants.}
    \label{fig:specw8}
\end{figure}

\textit{\color{blue}Single-band limit}. We begin by writing the Hamiltonian describing the dynamics of electrons in a crystal in the presence of a time dependent but spatially uniform electric field in the length gauge~\cite{PhysRevB.52.14636,PhysRevB.61.5337}:

\begin{equation}\label{hammilt}
    \hat{H}_{nm} = \delta_{nm}\epsilon_n(\mathbf{k})+e\hat{\mathbf{r}}_{nm}\cdot\mathbf{E}(t),
\end{equation}
here $n,m$ are band indices and $\mathbf{k}$ crystal momentum, $\epsilon_n(\mathbf{k})$ is the band energy dispersion,  $\hat{\mathbf{r}}_{nm}=i\delta_{nm}\partial_\mathbf{k}+\hat{\mathbf{A}}_{nm}$ is the position operator in the Bloch basis, and $\hat{\mathbf{A}}_{nm}({\bf k})=i \langle u_{n{\bf k}}|\partial_{\bf k} |u_{m{\bf k}}\rangle$ is the non-Abelian Berry connection~\cite{Blount}.
Before solving the full multiband problem, we consider the special limit in which electric field is slowly varying in time and the low energy dynamics of electrons can be described by projecting the Hamiltonian in Eq.\eqref{hammilt} onto a single band, which will illuminate on the deeper meaning of the BCD. In this limit, the acceleration operator can be shown to contain two terms (see S.I.\ref{supA}) :

\begin{figure}[t]
    \centering
    \includegraphics[width=0.48\textwidth]{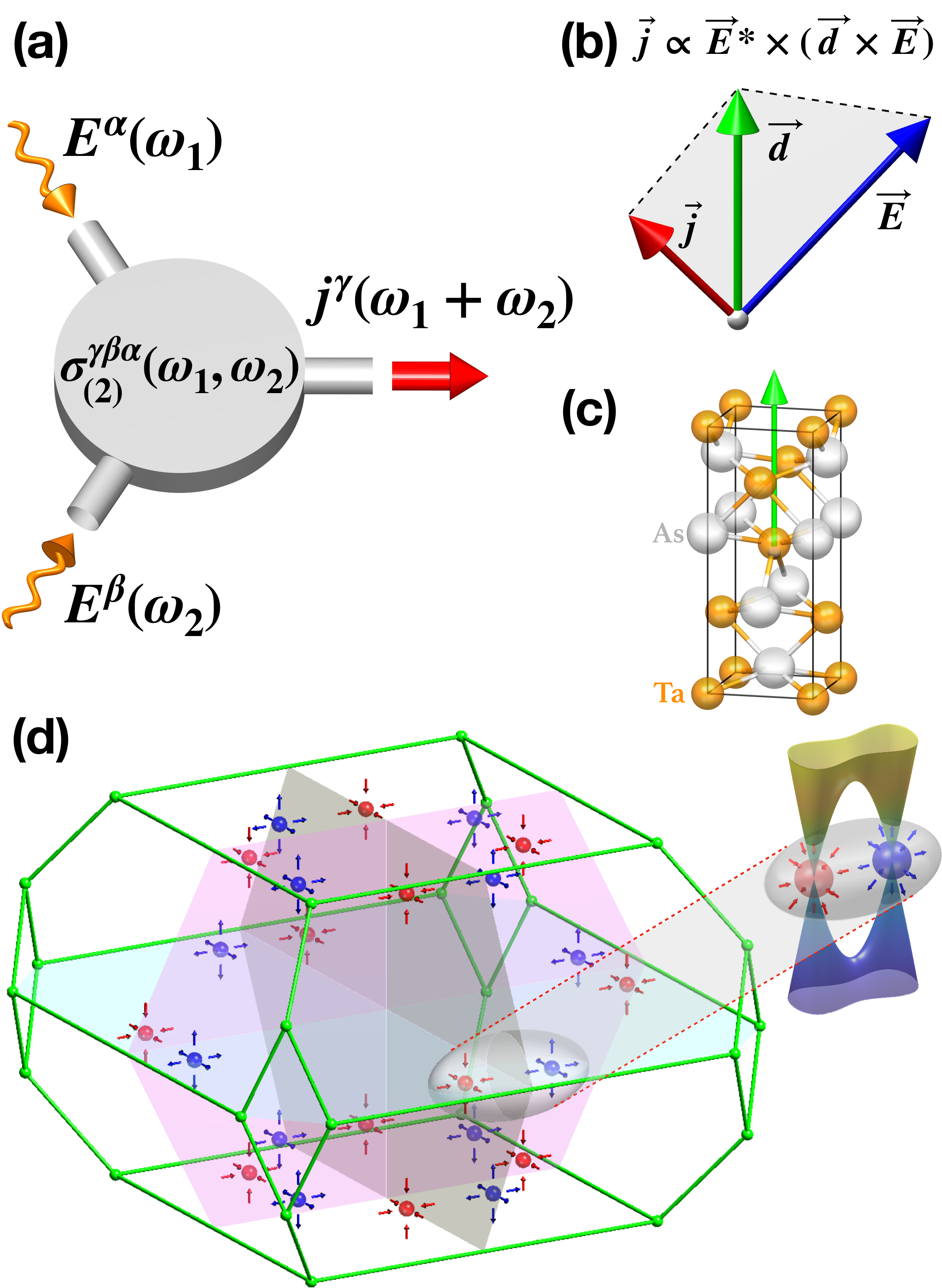}
    \caption{(a) Illustration of the second order conductivity tensor with two input frequencies, $\omega_{1,2}$, and space indices, $\alpha\beta$, for the driving electric field and one output current at frequency at $\omega_1+\omega_2$ with space index $\gamma$. (b) Depiction of the geometry of the BCD antisymmetric part~\cite{PhysRevLett.115.216806} (green vector), the driving linearly polarized electric (blue vector), and the second order electric current (red arrow). The three vectors are coplanar and the current is orthogonal to the electric field. (c) Crystal structure of TaAs with BCD (green arrow) oriented along its polar axis. (d) Weyl nodes in the Brillouin zone of TaAs with one elementary Weyl pair highlighted}
    \label{fig:sori}
\end{figure}

\begin{equation}\label{banda}
    \hat{a}^{\gamma}_{n} = -e\frac{E^{\alpha}}{\hbar^2}\frac{\partial^2}{\partial k^\alpha\partial k^\gamma}\epsilon_n+e^2\frac{E^\alpha E^\beta}{\hbar^2}\frac{\partial}{\partial k^\alpha} \hat{\Omega}_{n}^{\beta \gamma},
\end{equation}

here $\hat{\Omega}_{n}^{\alpha\beta}= \partial \hat{A}_{nn}^{\alpha}/\partial k^\beta-\partial \hat{A}_{nn}^{\beta}/\partial k^\alpha$ is the Abelian Berry curvature of the band “$n$”. The first term expresses the conventional Newton's second law and the tensor relating the electric field and the acceleration determines the ordinary linear Drude weight. The second term, however, is a non-Newtonian acceleration that is orthogonal to the applied electric field, which we call the $\textit{nonlinear Hall acceleration}$. The average over the occupied states of the tensor relating the nonlinear Hall acceleration and the bilinears of the electric field is precisely the BCD~\cite{PhysRevLett.115.216806} ($d^\lambda=\epsilon^{\lambda\alpha\beta}D^{\alpha\beta}=\epsilon^{\lambda\alpha\beta}\epsilon^{\beta\gamma\mu}(2\pi)^{-3}/2\sum_n\int d \mathbf{k}\Omega^{\mu\gamma}_{n}\partial f_n  /\partial k^\alpha$). Notice that there is no extrinsic quantity in this relation. Therefore the BCD can be interpreted as a nonlinear Drude weight, which is nonzero only when the system has a Fermi surface and breaks inversion symmetry.
In metals there is invariably a friction to the electron flow created by the impurities, phonons, and umklapp processes, that ultimately brings the liquid into a steady state with zero acceleration in the presence of an electric field. The terminal velocity will depend on the scattering rate from such agents, and this is why the nonlinear conductivity ends up depending on the scattering rate and not only on the BCD, in an analogous fashion to how the linear conductivity depends not only on the Drude weight but also on the scattering rate. Throughout our Letter we will take a simple relaxation time picture of disorder. The detailed role of disorder on the nonlinear conductivity of metals is indeed a subject of current intense investigation~\cite{2018arXiv181206989K,2018arXiv181208162I,2018arXiv181208377D,2019arXiv190104467N,2019arXiv190700577X}.

\textit{\color{blue}Multiband formalism}. We will now develop a multiband theory that is applicable to metals and insulators by modelling relaxation processes in a minimal fashion following the spirit of the relaxation time approximation, with the following non-Hermitian Liouville equation for the density matrix $\hat{\rho}$:

\begin{equation}\label{eq01}
    i\hbar\frac{d}{dt}\hat{\rho}(t)- [\hat{H}_0,\hat{\rho}(t)] =  -i\hbar\Gamma (\hat{\rho}(t)-\hat{\rho}^{0}),
\end{equation}
here $\Gamma$ is the relaxation rate and $\hat{\rho}_0$  is the equilibrium density matrix.
Following a standard perturbation theory analysis (see S.I.\ref{supB}) one obtains the full second order conductivity tensor with two input driving frequencies $\omega_{1,2}$  and the output current at a frequency $\omega_1 +\omega_2$  (see Fig.\ref{fig:sori}(a)), and which can be separated into metallic and interband terms as follows:

\begin{equation}\label{eq02}
    \sigma^{\gamma\beta\alpha}_{(2)}(\omega_1,\omega_2)=\sigma_{\mathrm{Met}}^{\gamma\beta\alpha}(\omega_1,\omega_2)+\sigma_{\mathrm{Inter}}^{\gamma\beta\alpha}(\omega_1,\omega_2).
\end{equation}

The interband terms contain the shift and injection currents identified in previous studies~\cite{PhysRevB.52.14636,PhysRevB.61.5337,PhysRevB.99.045121} and are reproduced in detail in S.I.\ref{supB}. The metallic terms are those which would manifestly vanish in the absence of a Fermi surface and that diverge in the low frequency $\omega_{1,2}\rightarrow 0$ and clean limits $\Gamma \rightarrow 0$. Their explicit expression is this:

\begin{multline}
    \sigma_{\mathrm{Met}}^{\gamma\beta\alpha}(\omega_1,\omega_2)=-\frac12\frac{e^3}{\hbar^2}\int\frac{d\mathbf{k}}{(2\pi)^3}\sum_{nm}\times\\\times\Bigg\{\frac{\frac{\partial \epsilon_n}{\partial k^\gamma}\frac{\partial^2 }{\partial^\alpha\partial^\beta}f_n\delta_{nm}}{(\omega_1+\omega_2+i\Gamma)(\omega_2+i\Gamma)}+\\+\frac{(\epsilon_m-\epsilon_n)\hat{A}^{\gamma}_{mn}\hat{A}^{\alpha}_{nm}\frac{\partial}{\partial k^\beta}(f_{m}-f_n)}{(\omega_{2}+i\Gamma)(\omega_1+\omega_2+\epsilon_m-\epsilon_n+i\Gamma)}+\\+\Bigg(\begin{array}{c}\alpha\leftrightarrow \beta\\\omega_1\leftrightarrow\omega_2 \end{array}\Bigg)\Bigg\},
\end{multline}
here $f_n$ is the Fermi Dirac distribution and ($\alpha\leftrightarrow \beta, \omega_1\leftrightarrow\omega_2$) denotes symmetrization under simultaneous swap of the indices $(\alpha,\beta)$ and the frequencies $(\omega_1,omega_2)$. The first term is a purely semiclassical Jerk term~\cite{PhysRevB.99.045121}. This term is distinct from the third order Jerk effect described in ~\cite{PhysRevLett.121.176604,2018arXiv180601206F} and vanishes under time-reversal-symmetric conditions which we will assume from here on. The second term is a multiband finite frequency generalization of the BCD nonlinear Hall conductivity, which at low frequencies reduces (see S.I.\ref{supB} Eq.(\ref{lfl})) to the expression in~\cite{PhysRevLett.115.216806}.

\begin{figure*}
\centering
\includegraphics[width=0.98\textwidth]{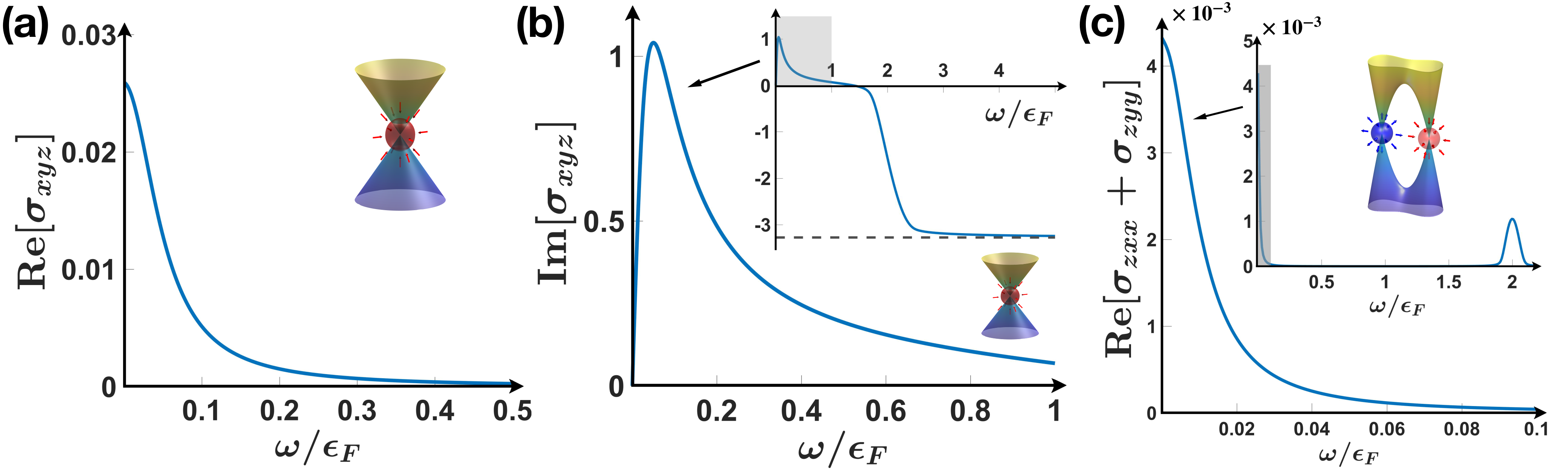}
\caption{Rectification conductivity (in units of $\{e^3/(\Gamma(2\pi)^3\hbar^2)\}$) for (a) linearly polarized light of a single Weyl node displaying the BCD Drude-like peak ($\hbar\Gamma = 0.05 \epsilon_F, u_z = 0.2v_0$). (b) circularly polarized light has two contributions: a BCD term at low frequency and an interband term (shown in the inset) which leads to the quantized circular photo-galvanic effect (dashed horizontal line) described in~\cite{PhysRevB.95.041104,deJuan2017}. (c) linearly polarized light of a Weyl node pair model relevant for TaAs, with its Drude-like BCD peak clearly separated from the interband term [displayed in inset, $\hbar\Gamma=0.01\epsilon_F,u_x = 0.05v_x, \Delta=5\epsilon_F, v_x=v_z=\sqrt{m/(2\Delta)}$]}
\label{fig3}
\end{figure*}

\textit{\color{blue}Quantum rectification sum rule}. We will state now one of the central findings of our study. Using the full expression for the second order conductivity, including the metallic and interband terms, and for any time reversal invariant material one can show (see S.I.\ref{supB}) that the following quantum rectification sum rule (QRSR) holds in the clean limit ($\Gamma \rightarrow 0$):
  
\begin{multline}\label{sumR}
    4\frac{\hbar^2}{e^3}\frac{1}{\pi}\int_{0}^{\infty} d\omega\mathrm{Re}\left[\sigma^{\gamma\beta\alpha}_{(2)}(-\omega,\omega)\right] =\\= \left\langle \frac{\partial}{\partial k^\beta}\hat{\Omega}^{\alpha\gamma}\right\rangle  
    +\left\langle[\hat{A}^{\beta},i\frac{\partial}{\partial k^\alpha} \hat{A}^{\gamma}]\right\rangle+\langle[\hat{A}^{\beta} ,[\hat{A}^\alpha,\hat{\bar{A}}^{\gamma}]]\rangle+\\+(\alpha\leftrightarrow \beta),
\end{multline}
where $\hat{\bar{A}}^{\alpha}_{nm} = \hat{{A}}^{\alpha}_{nm}(1-\delta_{nm})$ is the off-diagonal non-Abelian Berry connection and average is defined as follows:

\begin{equation}
    \langle \cdots \rangle = \sum_n \int \frac{d\mathbf{k}}{(2\pi)^d}f_n\langle n \mid(\cdots)\mid n\rangle.
\end{equation}

The integrand in this sum rule is the real part of the rectification conductivity which measures the net dc current produced by an ac linearly polarized electric field (see S.I.\ref{supC}). Remarkably, all the low frequency subgap spectral weight is exhausted by a delta function peak at zero frequency whose weight is given by the BCD, and which gives rise to the first term in the right-hand side of Eq.\eqref{sumR} (see Fig.\ref{fig:specw8}). The remaining weight accounting for the two other terms in the right-hand side arise from interband terms that are only nonzero when the frequency is above the spectral gap (see~Fig.\ref{fig:specw8}). It is remarkable that both intra- and interband terms integrate to a quantity that is purely quantum geometric depending only on the Berry connection. 
This sum rule offers a direct way to measure the BCD that bypasses knowledge of the disorder relaxation rate, by integrating the rectified current over frequency, in complete analogy to how the Drude weight is estimated from the conductivity in metals. A related rectification sum rule derived under more restricted conditions was recently reported in~\cite{PhysRevB.98.165113}.

This sum rule also opens a door to a systematic search for alternatives to conventional photovoltaics in which the band geometry plays a central role in the rectification mechanism \cite{RevModPhys.66.899,PhysRevB.98.165113}. This is because the rectification weight, defined as the right hand side of Eq.\eqref{sumR}, provides a natural figure of merit for photocurrent generation when the light spectrum is broad in comparison to the bandwidth of interest and its calculation could be efficiently streamlined with realistic modeling of band structures.
The rectification weight has units $L^{d-3}$, and, interestingly, it is dimensionless in 3D. The rectification weight of time reversal invariant materials therefore joins a select group of observables that are entirely controlled by the Berry geometry including the Hall conductivity~\cite{RevModPhys.82.1959,RevModPhys.82.1539}, the polarization~\cite{PhysRevB.47.1651,RevModPhys.66.899,RevModPhys.84.1419} and the magnetoelectric coefficient of 3D topological insulators~\cite{PhysRevB.81.159901,PhysRevB.81.205104}.

\textit{\color{blue}Weyl semimetals}. We will now apply our theory to Weyl semimetals which are topological states that can be realized in three-dimensional materials that break either inversion or time reversal symmetry~\cite{RevModPhys.90.015001}. Their nonlinear optoelectronic properties are a subject of intense current investigation~\cite{PhysRevB.95.041104,deJuan2017,Ma2017,2019arXiv190502236J,conv}. We will consider first the ideal model of an isolated Weyl node and subsequently a model of mirror related Weyl node pairs that captures the essential behavior of materials such as TaAs. A Hamiltonian for a single Weyl node is~\cite{RevModPhys.90.015001}:

\begin{equation}
    \hat{H}_0 = v_0\sum_{\alpha=x,y,z}{k_\alpha}\cdot{\hat{\sigma}_\alpha}+u_z k_z\hat{\mathds{1}}.
\end{equation}

Here $v_0$ is a Fermi velocity, $u_z$ is the tilt in $z$ direction, $\hat{\sigma}_\alpha$ are Pauli matrices and $\hat{\mathds{1}}$ is identity operator. In the clean limit, we have found that the rectification conductivity of an ideal Weyl node has only two contributions: the intraband BCD term and the interband injection current term responsible for the circular photogalvanic effect in \cite{PhysRevB.95.041104,deJuan2017}, as shown in (see~Fig.\ref{fig3}(a)-\ref{fig3}(c)). The BCD tensor of a Weyl node is symmetric and its nonzero components are

\begin{gather}\label{BCDres}
    (\mathrm{\hat{D}})^{\alpha\beta}=D^{\alpha\beta} =\frac12\varepsilon^{\beta \gamma\mu}\sum_n \int \frac{d\mathbf{k}}{(2\pi)^3}\hat{\Omega}^{\mu\gamma}_n\frac{\partial}{\partial k^\alpha} f_n,\\
    D^{xx}=D^{yy} =\frac{1}{8\pi^2}\frac{1}{\tilde{u}^3}\left(\tilde{u}+\frac{\tilde{u}^2-1}{2}\ln\frac{1+\tilde{u}}{1-\tilde{u}}\right),\\ D^{zz}=\frac{1}{4\pi^2}\frac{\tilde{u}^2-1}{\tilde{u}^3}\left(\tilde{u}-\frac12\ln\frac{1+\tilde{u}}{1-\tilde{u}}\right),
\end{gather}

here $\tilde{u} ={u_z}/{v_0}$. Notice that the BCD of a single Weyl point does not depend on the location of the Weyl point in momentum space, as found in \cite{PhysRevB.97.195151}. The origin of this inconsistency is discussed in the supplementary (see S.I.\ref{secD}). Remarkably the trace of BCD for a Weyl node is a universal quantity independent of the details of its Hamiltonian determined by its integer valued monopole strength  ($\mathcal{S}_j \in \mathbb{Z}$)~\cite{RevModPhys.90.015001}:

\begin{equation}
    \mathrm{Tr[\hat{D}]} =  \frac{\mathcal{S}_j}{4\pi^2}.
\end{equation}

The BCD rectification conductivity is given by

\begin{multline}    
\mathbf{j}(0)_{\mathrm{BD}} = \frac{e^3}{\hbar^2}\int_{-\infty}^{\infty} d\omega \frac{i}{\omega+i\Gamma}\Big[\left\{\hat{\mathrm{D}}.\mathbf{E}(-\omega)\right\}\times \mathbf{E}(\omega)\Big].
\end{multline}

Therefore the QRSR for an ideal Weyl node yields 

\begin{equation}
    \int_{-\infty}^{\infty} \sigma^{\gamma\beta\alpha}_{(2)}(-\omega,\omega)d\omega=\frac{e^3}{\hbar^2}\frac{D^{zz}-D^{xx}}{2}\left\{\delta^{\alpha z}\varepsilon^{\gamma z\beta}+(\alpha\leftrightarrow \beta)\right\}.
\end{equation}

Notice that Weyl nodes related by time reversal symmetry have the same BCD tensors, and, therefore, even though the responses from Weyl
nodes of opposite topological charge tend to cancel, the net response in materials without symmetries relating nodes of opposite charge, such as SrSi2~\cite{Huang2016,Chang2018TopologicalQP}, will be finite.
However, several Weyl materials such as TaAs~\cite{Huang2015,PhysRevX.5.011029,Xu2015,PhysRevX.5.031013} have mirror symmetries mapping Weyl nodes of opposite charge. In fact, for TaAs with space group I41md only the antisymmetric part of the BCD tensor is allowed \cite{PhysRevLett.115.216806}, and therefore, the contribution to the net BCD tensor from the linearized model cancels after adding all nodes. However, TaAs and related materials are expected to have a large BCD~\cite{PhysRevB.97.041101}. A reason for this enhancement in TaAs is the proximity of pairs of nodes of opposite charge (see Fig.\ref{fig:sori}(d)). Related large enhancements have been predicted in BiTeI~\cite{PhysRevLett.121.246403}. The ideal Hamiltonian describing such pairs is~\cite{RevModPhys.90.015001,PhysRevLett.121.246403}

\begin{equation}\label{DwHamm}
    \hat{H} = v_xk_x\hat{\sigma}_x+\frac{\lambda-k_y^2}{2m}\hat{\sigma}_y + v_zk_z\hat{\sigma}_z+u_xk_x\hat{\mathds{1}}.
\end{equation}

Here $v_x,v_y = \sqrt{|\lambda|}/m,v_z$ are anisotropic Fermi velocities, $u_x$ is tilt in the $x$ direction and $2\lambda$ is the node shift. The antisymmetric BCD tensor (defined as: $d^\alpha=\epsilon^{\alpha\beta\gamma} D^{\beta\gamma}$) is a vector oriented along the polar axis of TaAs (see~Fig.\ref{fig:sori}(b) and Fig.\ref{fig:sori}(c)), and the contribution from each Weyl-nodepair to the BCD is

\begin{equation}
    d_{z}\approx-\frac{3v_x v_z u_x n^{1/3}}{10\pi^{4/3}v_y^2m(3v_xv_yv_z)^{2/3}},
\end{equation}
 where $n$ is a density of carriers.
The rectification response to linearly polarized fields, shown in (see~Fig.\ref{fig3}(c)), displays a clear separation between the interband and the BCD term, making it viable to estimate the BCD by integrating it up to some $\omega\ll 2 \epsilon_F$.

Fig.\ref{fig3} shows the response of these ideal models for Weyl nodes. The real and imaginary parts of the second order conductivity have a crucially distinct nature, because the former controls the response to time-reversal-symmetric drive (linearly polarized light) while the second to time-reversal-breaking drive (circularly polarized light)(see S.I.\ref{supC} Eq.(\ref{currre})). The QRSR requires knowledge only of the real part. Figure (a) illustrates this for a single Weyl node. Such contribution would cancel when adding Weyl pairs related by a mirror symmetry as it is the case for TaAs. Figure (c), however, illustrates that there is a finite contribution beyond the linearized model even in the case of Weyl pairs related by a mirror, which could be observed in TaAs. For completeness we also show the response to circularly polarized light in figure (b), illustrating how our formalism interpolates from the interband effects of~\cite{PhysRevB.95.041104,deJuan2017} to the low frequency intraband BCD peak.

\textit{\color{blue}Discussion}. We have shown that the BCD controls a nonlinear Hall acceleration in metals without inversion symmetry that scales with the square of the applied electric field. Therefore, the BCD can be viewed as a nonlinear version of the Drude weight, that serves as a measurable order parameter for metallic inversion symmetry breaking. 

We have also shown that the nonlinear conductivity of time-reversal invariant materials satisfies a QRSR whose intraband weight is exhausted by a sharp Drude-like BCD peak, and, remarkably, also the inter-band contributions to this QRSR are purely controlled by the Berry connection, with all dependence on band energies disappearing. The rectification weight defined by this sum rule, therefore, provides a figure of merit for dc current generation in response to light, provided that the incident spectrum is broad in comparison to the band energy window of interest. We hope that this rectification weight could be estimated from first principle studies to help guide the search for alternative photovoltaic materials based on the quantum geometry of Bloch bands. Another promising avenue of applications for these effects are wireless energy harvesting devices~\cite{2018arXiv181208162I,2019arXiv190202514H}.

Weyl semimetals are also promising platforms to study these effects. In particular, several polar Weyl materials such as TaAs will have a large vectorial component of the BCD aligned with its polar axis (see Fig.\ref{fig:sori}(b) and Fig.\ref{fig:sori}(c)). The BCD and the QRSR in these materials could be studied by measuring a nonlinear Hall current flowing along the polar axis in response to a driving electric field in the plane perpendicular to the polar axis as a function of frequency. 

Cold atomic systems also offer an interesting alternative platform to investigate these nonlinear effects of the Berry geometry~\cite{2015NatPh.11.162A,jotzuexperimental2014,Flaschner1091,RevModPhys.89.011004,2019NatPh..15..449A,2015Sci...347..288D},
 where the slower time scales for dynamics and the absence of friction might allow us to directly measure the nonlinear Hall acceleration.

We are thankful to Liang Fu for previous collaborations
and discussions that inspired this work. We thank Fernando
de Juan and Adolfo Grushin for valuable discussions, and
for sharing their manuscript on a study employing a similar
formalism prior to publication~\cite{conv}, and Elio König, Habib
Rostami, Roser Valenti, and Christof Weitenberg for
valuable discussions and correspondence.

\vskip 0.4in
\section{Supplementary Information}
\subsection{Nonlinear Hall acceleration} \label{supA}
Following Bloch's theorem, electronic states in crystals are labelled by a crystal momentum ${\bf k}$ which belongs to a periodic Brillouin zone and a discrete band index $n$. We express the position operator in such Bloch basis, which, as first demonstrated by Blount~\cite{Blount}, is comprised of two pieces:

\begin{equation}
     \mathbf{\hat{r}}_{nm}=\delta_{nm} i\partial_{\bf k}+\hat{\mathbf{A}}_{nm}({\bf k}),
\end{equation}
where $\hat{\mathbf{A}}_{nm}({\bf k})$ is the non-Abelian Berry connection which is related to the periodic part of the Bloch states, $|u_{m{\bf k}}\rangle$, as $\hat{\mathbf{A}}_{nm}({\bf k})=i \langle u_{n{\bf k}}|\partial_{\bf k} |u_{m{\bf k}}\rangle$.
With this representation the Hamiltonian of the electron in the presence of a uniform electric field reads as:

\begin{equation}\label{hammilt2}
    \hat{H} = \hat{H}_0(\mathbf{k})+e\hat{\mathbf{r}}\cdot\mathbf{E}(t),
\end{equation}

We now project this Hamiltonian onto a single band $n$. A consequence of this is that the projected position $\hat{\mathbf{r}}_{n}=\hat{\mathbf{r}}_{nn}$ does not commute with itself along different directions. The commutator is the Abelian Berry curvature of such band:

\begin{equation}
    [\hat{r}^\alpha_n,\hat{r}^\beta_n] = i\left(\frac{\partial}{\partial{k^{\alpha}}}\hat{A}^{\beta}_n-\frac{\partial}{\partial{k^{\beta}}}\hat{A}^{\alpha}_n\right) = i\hat{\Omega}_{n}^{\alpha\beta}.
\end{equation}

Here $\mathbf{A}_n=\mathbf{A}_{nn}$.From this one can readily derive the velocity operator that contains the well-known anomalous velocity~\cite{RevModPhys.82.1959,RevModPhys.82.1539}: 

\begin{equation}\label{veloc}
    \hat{v}^{\alpha}_n = \frac{i}{\hbar}[\hat{H}(\mathbf{k}),\hat{r}^\alpha]_n=\frac{1}{\hbar}\frac{\partial}{\partial k^{\alpha}}\epsilon_n -e\frac{E^\beta(t)}{\hbar}\hat{\Omega}^{\beta\alpha}_n  .
\end{equation}

 Let us now consider the acceleration operator defined as the rate of change of the velocity. This operator can be found to be comprised of three pieces:

\begin{equation}\label{banda2}
    \hat{a}^{\gamma}_{n} = -e\frac{E^{\alpha}}{\hbar^2}\frac{\partial^2}{\partial k^\alpha\partial k^\gamma}\epsilon_n-e\frac{1}{\hbar}\frac{\partial E^\alpha}{\partial t}\hat{\Omega}^{\alpha\gamma }_{n}+e^2\frac{E^\alpha E^\beta}{\hbar^2}\frac{\partial}{\partial k^\alpha} \hat{\Omega}^{\beta \gamma}_{n}.
\end{equation}

The first piece is Newton's second law in which the acceleration is proportional to the force times the second derivative of the dispersion with respect to momentum, which measures the inverse inertia or Drude weight. The second piece is proportional to the derivative of the electric field is related to the instantaneous change of the anomalous velocity described in Eq.\eqref{veloc} and is negligible when the field varies sufficiently slowly in time. The third piece is the non-Newtonian acceleration that is quadratic in the applied electric field and is proportional to the Berry curvature dipole: the gradient in momentum space of the Berry curvature~\cite{PhysRevLett.115.216806}. This term is the one that we refer to as Nonlinear Hall acceleration throughout the main text. We also note that the gradient of the Berry curvature is even under time-reversal but odd under space inversion operations.
\subsection{Full expression for second order conductivity} \label{supB}

Following perturbation theory:

    \begin{equation}
    \langle j ^\gamma(t) \rangle_2=\int_{-\infty}^{\infty}\int_{-\infty}^{\infty}dt_1 dt_2 E^{\beta}(t_2)E^{\alpha}(t_1)\sigma^{\gamma\beta\alpha}_{(2)}(t,t_1,t_2) ,
    \end{equation}
    
    based on a Liouville Eq.(\ref{eq01}) for density matrix:
    
    \begin{equation}
        \hat{\rho}(t)=\hat{\rho}_0+e^{-\Gamma t}\delta\hat{\rho}(t),\qquad
        \langle n|\hat{\rho}_0|m\rangle =f_n\delta_{nm} ,
    \end{equation}
    we induced nontrivial relaxation $\Gamma$. According to this approach second order conductivity is written as:
    
    \begin{equation}\label{tcond}
    \sigma^{\gamma\beta\alpha}_{(2)}(t,t_1,t_2) =-\frac{e^3}{\hbar^2}e^{\Gamma(t_2-t)}\langle[r^\beta(t_2),[r^\alpha(t_1),v^{\gamma}(t)]]\rangle,
    \end{equation}

for $t\geq t_1\geq t_2$ and $0$ otherwise. 

In frequency domain, the second-order conductivity can be expressed as a function of two driving frequencies with the frequency of the physical current being the sum of these two, as depicted in Fig.\ref{fig:sori}(a). We have computed all the contributions to the second order current that would be present in a multiband system for any frequency of the driving fields. The conductivities can be separated into "metallic" and interband terms by isolating the parts that have physical divergences in the limit in which both driving frequencies vanish $\omega\rightarrow0$, and those that do not, as described in Eq.(\ref{eq02}) of the main text.

We have encountered two metallic terms in the nonlinear conductivity that can be termed the semiclassical jerk (SCJ) and berry curvature dipole terms, and whose expressions are:  

\begin{equation}\label{fullcond}
    \sigma^{\gamma\beta\alpha}_{\mathrm{Met}}(\omega_1,\omega_2) =\sigma^{\gamma\beta\alpha}_{\mathrm{SCJ}}(\omega_1,\omega_2)+\sigma^{\gamma\beta\alpha}_{\mathrm{BCD}}(\omega_1,\omega_2),
\end{equation}   
\begin{multline}
    \sigma^{\gamma\beta\alpha}_{\mathrm{SCJ}}(\omega_1,\omega_2) = -\frac12\frac{e^3}{\hbar^2}\int\frac{d\mathbf{k}}{(2\pi)^3}\times\\\times\sum_{nm}\frac{\frac{\partial \epsilon_n}{\partial k^\gamma}\frac{\partial^2 }{\partial^\alpha\partial^\beta}f_n\delta_{nm}}{(\omega_1+\omega_2+i\Gamma)(\omega_2+i\Gamma)}+\left(\begin{array}{c}\alpha\leftrightarrow \beta\\\omega_1\leftrightarrow\omega_2\end{array}\right),
\end{multline}    
\begin{multline}    
    \sigma^{\gamma\beta\alpha}_{\mathrm{BCD}}(\omega_1,\omega_2) =-\frac12\frac{e^3}{\hbar^2}\int\frac{d\mathbf{k}}{(2\pi)^3}\times\\\times\sum_{nm}\frac{(\epsilon_m-\epsilon_n)\hat{A}^{\gamma}_{mn}\hat{A}^{\alpha}_{nm}\frac{\partial}{\partial k^\beta}(f_{m}-f_n)}{(\omega_{2}+i\Gamma)(\omega_1+\omega_2+\epsilon_m-\epsilon_n+i\Gamma)}+\\+\left(\begin{array}{c}\alpha\leftrightarrow \beta\\\omega_1\leftrightarrow\omega_2\end{array}\right).
\end{multline}

The semiclassical jerk term is distinct from the third order jerk effect described in~\cite{PhysRevLett.121.176604,2018arXiv180601206F}.
These terms remain finite in the expressions $\omega\rightarrow0$ only thanks to the fact that we have accounted for a finite relaxation rate. The jerk term identically vanishes for a time reversal invariant system. The term that we have called the Berry curvature dipole reduces to the semiclassical result obtained in \cite{PhysRevLett.115.216806} in the limit in which the external frequency of the current is much smaller that the interband optical threshold:

\begin{multline}\label{lfl}
    \sigma^{\gamma\beta\alpha}_{\mathrm{BCD}}(\omega_1+\omega_2\ll \epsilon_{nm},\omega_1,\omega_2)=\\=-\frac12\frac{e^3}{\hbar^2}\frac{i}{\omega_{2}+i\Gamma}\int\frac{d\mathbf{k}}{(2\pi)^3}\sum_{n}\hat{\Omega}^{\gamma \alpha}_{n}\frac{\partial}{\partial k^\beta}f_n+\\+\left(\begin{array}{c}\alpha\leftrightarrow \beta\\\omega_1\leftrightarrow\omega_2\end{array}\right).
\end{multline}

In addition to these terms we have the interband terms defined as those that remain finite when both driving frequencies vanish. These interband terms contain the shift and injection currents obtained in previous work \cite{PhysRevB.52.14636,PhysRevB.61.5337,PhysRevB.99.045121}, and can be explicitly written as:

\begin{equation}\label{conductiv2}
    \sigma^{\gamma\beta\alpha}_{\mathrm{Inter}}(\omega_1,\omega_2) = \sigma^{\gamma\beta\alpha}_{\mathrm{I}}(\omega_1,\omega_2)+\sigma^{\gamma\beta\alpha}_{\mathrm{IB2}}(\omega_1,\omega_2),
\end{equation}
\begin{equation}\label{conductiv}
 \sigma^{\gamma\beta\alpha}(\omega_1,\omega_2)=\sigma^{\gamma\beta\alpha}_{\mathrm{Met}}(\omega_1,\omega_2)+\sigma^{\gamma\beta\alpha}_{\mathrm{Inter}}(\omega_1,\omega_2),
\end{equation}
we labeled interband terms as: "Injection" and "Inter-band 2" parts, which explicitly read as:

\begin{multline}
    \sigma^{\gamma\beta\alpha}_{\mathrm{I}}(\omega_1,\omega_2) = \frac12\frac{e^3}{\hbar^2}\left(1+\frac{i\Gamma}{\omega_1+\omega_2+i\Gamma}\right)\int\frac{d\mathbf{k}}{(2\pi)^3}\times\\\times\sum_{nm}\frac{(f_{n}-f_m)\hat{A}^{\beta}_{nm}\hat{A}^{\alpha}_{mn}(\frac{\partial}{\partial k^\gamma}\epsilon_n-\frac{\partial}{\partial k^\gamma}\epsilon_m)}{(\omega_1+\epsilon_n-\epsilon_m+i\Gamma)(\omega_2-\epsilon_n+\epsilon_m+i\Gamma)}+\\+\left(\begin{array}{c}\alpha\leftrightarrow \beta\\\omega_1\leftrightarrow\omega_2\end{array}\right),
\end{multline}
\begin{multline}
    \sigma^{\gamma\beta\alpha}_{\mathrm{IB2}}(\omega_1,\omega_2) = \frac12\frac{e^3}{\hbar^2}\int\frac{d\mathbf{k}}{(2\pi)^3}\times\\\times\sum_{nm}\Bigg\{ \frac{ \hat{A}^{\gamma}_{mn}(\epsilon_{m}-\epsilon_n)}{\omega_1+\omega_2-\epsilon_{n}+\epsilon_{m}+i\Gamma}\frac{\partial}{\partial k^\alpha}\frac{(f_{n}-f_m)\hat{A}^{\beta}_{nm}}{\omega_2-\epsilon_{n}+\epsilon_{m}+i\Gamma} 
    +\\+ i\frac{(f_{n}-f_m)\hat{A}^{\beta}_{nm}}{\omega_2  -\epsilon_{n}+\epsilon_{m}+i\Gamma}\sum_c \bigg[\frac{\hat{A}^\alpha_{mc}\hat{A}^{\gamma}_{cn}(\epsilon_c-\epsilon_n)}{\omega_1+\omega_2 - \epsilon_{n}+\epsilon_c+i\Gamma}-\\-\frac{\hat{A}^{\gamma}_{mc}\hat{A}^\alpha_{cn}(\epsilon_m-\epsilon_c)}{\omega_1+\omega_2 - \epsilon_{c}+\epsilon_m+i\Gamma}\bigg]\Bigg\}+\left(\begin{array}{c}\alpha\leftrightarrow \beta\\\omega_1\leftrightarrow\omega_2\end{array}\right).
\end{multline}
The second order conductivity which we introduced above has the following property:

\begin{equation}
    \sigma^{\gamma\beta\alpha}_{(2)}(\omega_1,\omega_2)^{*}=\sigma^{\gamma\beta\alpha}_{(2)}(-\omega_1,-\omega_2).
\end{equation}

  The theory here described therefore unifies a large class of nonlinear optoelectronic phenomena in insulators and metals, including rectification and second harmonic generation processes.

  \subsection{Quantum rectification sum rule}
  Here we will describe the sum rule for the rectification conductivity that appears in time reversal invariant systems in the limit in which the relaxation rate $\Gamma$ is smaller than the optical $\epsilon_{n}-\epsilon_m$ separating occupied and empty states. 

To obtain rectification conductivity in Eq.\eqref{conductiv} we set $\omega_1+\omega_2 = 0$, and integrate over one driving frequency as follows:  

\begin{multline}
   \frac2\pi \int_{-\infty}^{\infty}  \sigma^{\gamma\beta\alpha}_{(2)}(-\omega,\omega)d\omega =  -\frac{e^3}{\hbar^2}\int\frac{d\mathbf{k}}{(2\pi)^3}\times\\\times\sum_n\Bigg\{\frac{1}{\Gamma}\bigg(-\frac{\partial}{ \partial k^{\gamma}}\epsilon_{n}\frac{\partial^2}{\partial k^\alpha \partial k^\beta }f_n+\\+\sum_m(f_{n}-f_m)\hat{A}^{\beta}_{nm}\hat{A}^{\alpha}_{mn}(\frac{\partial}{\partial k^{\gamma}}\epsilon_{n}-\frac{\partial}{\partial k^{\gamma}}\epsilon_{m})\bigg)+\\+i\sum_{m}\bigg[\bigg(\frac{\hat{A}^\alpha_{nm}\hat{A}^{\gamma}_{mn}(\epsilon_{m}-\epsilon_n)}{ \epsilon_{m}-\epsilon_n+i\Gamma}\bigg)\frac{\partial}{\partial k^\beta} (f_{n}-f_m)+\\+ \frac{\hat{A}^{\gamma}_{mn}(\epsilon_{m}-\epsilon_n)}{\epsilon_{m}-\epsilon_n+i\Gamma}\frac{\partial}{\partial k^\alpha} (f_{n}-f_m)\hat{A}^{\beta}_{nm}\bigg ]
    -\\-\sum_m (f_{n}-f_m)\hat{A}^{\beta}_{nm}\sum_c \bigg[\frac{\hat{A}^\alpha_{mc}\hat{\bar{A}}^{\gamma}_{cn}(\epsilon_{c}-\epsilon_n)}{ \epsilon_{c}-\epsilon_n+i\Gamma}-\\-\frac{\hat{\bar{A}}^{\gamma}_{mc}\hat{A}^\alpha_{cn}(\epsilon_{m}-\epsilon_c)}{ \epsilon_{m}-\epsilon_c+i\Gamma}\bigg]\Bigg\}+(\alpha\leftrightarrow \beta).
\end{multline}

 Now we take the clean limit ($\Gamma\ll\epsilon_{n}-\epsilon_{m}, \forall n,m$) and assume time-reversal symmetry. With time reversal symmetry we have the following property:

\begin{gather}
    \mathcal{T}\hat{\mathbf{r}}\mathcal{T}^{-1}=\hat{\mathbf{r}},\\
    \mathcal{T} \hat{\mathbf{v}}\mathcal{T}^{-1} = -\hat{\mathbf{v}}.
\end{gather}



According to this symmetry and after taking the clean limit one obtains:

\begin{multline}\label{sumR2}
    4\frac{\hbar^2}{e^3}\frac{1}{\pi}\int_{0}^{\infty} d\omega\mathrm{Re}\left[\sigma^{\gamma\beta\alpha}_{(2)}(-\omega,\omega)\right] =\\= \left\langle \frac{\partial}{\partial k^\beta}\hat{\Omega}^{\alpha\gamma}\right\rangle  
    +i\left\langle[\hat{A}^{\beta},\frac{\partial}{\partial k^\alpha} \hat{A}^{\gamma}]\right\rangle+\langle[\hat{A}^{\beta} ,[\hat{A}^\alpha,\hat{\bar{A}}^{\gamma}]\rangle+\\+(\alpha\leftrightarrow \beta).
\end{multline}

This sum rule contains information only about Berry connection and it is therefore a purely quantum geometrical property of the material.

\subsection{Rectification for linearly polarized light}\label{supC}
In this section we will show that linearly polarized light produces the rectification current which is controlled by real part of the second order conductivity. 
Consider a monocromatic driving electric field at frequency omega:

\begin{equation}
    E_{\alpha}(t) = \int d\omega E_\alpha(\omega)e^{i\omega t}=E_\alpha e^{i\omega_0t}+E^{*}_\alpha e^{-i\omega_0t},
\end{equation}

the rectification current is:

\begin{multline}\label{currre}
    j^{\gamma}(0)=\sigma_{(2)}^{\gamma\beta\alpha}(\omega_0,-\omega_0)E_{\alpha}E_{\beta}^*+\sigma_{(2)}^{\gamma\beta\alpha }(-\omega_0,\omega_0)E_{\alpha}^*E_{\beta}=\\=\sigma_{(2)}^{\gamma\beta\alpha }(\omega_0,-\omega_0)E_{\alpha}E_{\beta}^*+\sigma_{(2)}^{*\gamma\beta\alpha }(\omega_0,-\omega_0)E_{\alpha}^*E_{\beta}=\\=\mathrm{Re}\left[\sigma_{(2)}^{\gamma\beta\alpha}(\omega_0,-\omega_0)\right](E_{\alpha}E_{\beta}^*+E_{\alpha}^*E_{\beta})+\\+\mathrm{Im}\left[\sigma_{(2)}^{\gamma\beta\alpha }(\omega_0,-\omega_0)\right](E_{\alpha}E_{\beta}^*-E_{\alpha}^*E_{\beta}).
\end{multline}

For linearly polarized light all the components can be taken to have a common complex phase, as follows:

\begin{equation}
    E_\alpha=\mathcal{E}_\alpha e^{i\Phi},
\end{equation}
where $\mathcal{E}_\alpha$ is real. Finally the rectification current is:

\begin{equation}
    j^{\gamma}(0)=2\mathrm{Re}\left[\sigma_{(2)}^{\gamma\beta\alpha }(\omega_0,-\omega_0)\right]\mathcal{E}_{\alpha}\mathcal{E}_{\beta}.
\end{equation}

Therefore we see that the rectification conductivity for linearly polarized light is controlled by the Real part of the second order conductivity.

We will now explain why the quantum rectification sum rule can be viewed as a figure of merit for the current generation of a bulk material in an ideal case in which the incident light has a spectral distribution which is broader than the bandwidth of interest. The incident light will have a certain spectral intensity distribution. For example, in the case of sunlight this can be approximated as the distribution from the black-body radiation with the temperature of the sun. Because the incident light is random, there will be equal  probability to have right-circular and left-circular polarized light and therefore only the response to linearly polarized light will be important in average, because for a time reversal invariant material the current response to left and right-circularly polarized light has exactly the same magnitude but opposite directions. Therefore, assuming that the incident light is isotropic and characterized by an spectral intensity distribution $I(\omega)$, the net current rectified will be given by:

\begin{equation}\label{photocurr}
    J^{\gamma} = \sum_\alpha\int d\omega I(\omega)\mathrm{Re}\left[\sigma_{(2)}^{\gamma\alpha\alpha}(\omega,-\omega)\right]
\end{equation}

Now let us suppose that we would like to characterize the ability to generate current in response to light for a certain specific group of bands in a material. We can imagine fitting a tight-binding model for such bands. In such case, the integrand in Eq.(\ref{photocurr}) would only be non-zero up to a maximum frequency given by the bandwidth of such tight-binding model. Now, assuming that the spectral intensity distribution of incident light does not change substantially over the bandwidth of the bands of interest, and can thus approximated as a constant over such window, $I(\omega)\approx I_0$, we can approximate the total rectified current as:

\begin{equation}
    J^{\gamma} = I_0\sum_\alpha\int d\omega \mathrm{Re}\left[\sigma_{(2)}^{\gamma\alpha\alpha}(\omega,-\omega)\right]
\end{equation}

And, therefore, using Eq.(\ref{sumR2}), we see that the net rectified current is controlled by the QRSR in the ideal limit of incident light with a spectral distribution that is much broader than the bandwidth of interest.

\vspace{7mm}
\subsection{BCD for Weyl models}\label{secD}
In this section we describe the derivation of BCD tensor for single and double Weyl node Hamiltonians. The single-Weyl Hamiltonian is:

\begin{equation}\label{generalSw}
    \hat{H}_0 = \sum_{\alpha=x,y,z}v_\alpha{(k_\alpha-b_\alpha)}{\hat{\sigma}_\alpha}+\sum_{\alpha=x,y,z}u_\alpha (k_\alpha-b_\alpha),
\end{equation}
where $\mathbf{b}$ is a constant vector which determines the location of the Weyl point in momentum, $v_\alpha$ are Fermi velocities and $u_\alpha$ are tilts in appropriate direction. According to Eq.\eqref{lfl} the BCD contribution to the rectification conductivity is:

\begin{multline}\label{bcdtocomp1}
    \sigma^{\gamma\beta\alpha}_{\mathrm{BCD,I}}(\omega_1+\omega_2\ll \epsilon_{nm},\omega_1,\omega_2)=\\=-\frac12\frac{e^3}{\hbar^2}\frac{i}{\omega_{2}+i\Gamma}\int\frac{d\mathbf{k}}{(2\pi)^3}\sum_{n}\hat{\Omega}^{\gamma \alpha}_{n}\frac{\partial}{\partial k^\beta}f_n+\\+\left(\begin{array}{c}\alpha\leftrightarrow \beta\\\omega_1\leftrightarrow\omega_2\end{array}\right),
\end{multline}

This expression makes manifest that at low temperatures only states near the Fermi surface contribute to the BCD conductivity. It also makes manifest that   at zero temperature this integral is non-divergent because it is restricted to a finite region of integration, provided that the Berry curvature has no singularities at the Fermi surface. Moreover, for the Weyl model in Eq.(\ref{photocurr}) we have that $f_n(k,b)= f_n(k-b,0)$, and $\Omega_n(k,b)=\Omega_n(k-b,0)$, and therefore it is clear that the BCD conductivity is independent of $b$. This also implies that that the BCD conductiivity of a collection of Weyl points is independent of their particular locations and relative distances in momentum space, provided that they  can be described by a sum of decoupled Hamiltonians of the form Eq.\eqref{generalSw} with different $b$ for each of the Weyl nodes.

Let us now contrast this formula for the BCD with that in which the derivatives act on the Berry curvature rather than the distribution function, which is given by:

\begin{multline}\label{bcdtocomp2}
     \sigma^{\gamma\beta\alpha}_{\mathrm{BCD,II}}(\omega_1+\omega_2\ll \epsilon_{nm},\omega_1,\omega_2)=\\=\frac12\frac{e^3}{\hbar^2}\frac{i}{\omega_{2}+i\Gamma}\int\frac{d\mathbf{k}}{(2\pi)^3}\sum_{n}f_n\frac{\partial}{\partial k^\beta}\hat{\Omega}^{\gamma \alpha}_{n}+\\+\left(\begin{array}{c}\alpha\leftrightarrow \beta\\\omega_1\leftrightarrow\omega_2\end{array}\right),
\end{multline}
sub-index II means that this is alternative way of BCD computing. Their difference is a total derivative given by:

\begin{multline}\label{bcdtocomp3}
    \sigma^{\gamma\beta\alpha}_{\mathrm{BCD},\mathrm{I}}(\omega_1,\omega_2)-\sigma^{\gamma\beta\alpha}_{\mathrm{BCD},\mathrm{II}}(\omega_1,\omega_2)=\\=-\frac12\frac{e^3}{\hbar^2}\frac{i}{\omega_{2}+i\Gamma}\int\frac{d\mathbf{k}}{(2\pi)^3}\sum_{n}\frac{\partial}{\partial k^\beta}\left(f_n\hat{\Omega}^{\gamma \alpha}_{n}\right)+\\+\left(\begin{array}{c}\alpha\leftrightarrow \beta\\\omega_1\leftrightarrow\omega_2\end{array}\right).
\end{multline}

Therefore if the Hamiltonian describes a proper lattice model both definitions are equivalent because the periodicity of the Brillouin zone would enforce the intregral of the total derivative to vanish. However, for a continuum effective Hamiltonian the two definitions could differ, and, when they do, the correct one should be taken to be $\sigma^{\gamma\beta\alpha}_{\mathrm{BCD},\mathrm{I}}(\omega_1,\omega_2)$. In particular, in the case of the single Weyl node described by Eq.\eqref{generalSw} we see that the fully occupied negative energy states contribute to the integral in Eq.\eqref{bcdtocomp3}, and therefore this integral can approach a constant  as the UV cutoff, $K_{UV}$, is send to infinity because the surface area grows as $K_{UV}^2$ , while the Berry curvature decreases as $1/K_{UV}^2$. Therefore, one can verify that computing the BCD conductivity from $\sigma^{\gamma\beta\alpha}_{\mathrm{BCD},\mathrm{II}}(\omega_1,\omega_2)$ leads to an artefact according to which the BCD
has a nontrivial dependence on the location of the origin of the Weyl nodes~\cite{PhysRevB.97.195151}.







\bibliography{mysuperbib}

\end{document}